\documentclass{article}

\usepackage{arxiv}
\usepackage{amsthm}

\newtheoremstyle{bolddefinition}
  {3pt} 
  {3pt} 
  {\itshape} 
  {} 
  {\bfseries} 
  {.} 
  {.5em} 
  {} 

\theoremstyle{bolddefinition}
\newtheorem{definition}{Definition}[section]
\usepackage{colortbl} 
\usepackage{xcolor} 
\usepackage{graphicx} 
\usepackage{algorithm}
\usepackage{algpseudocode}
\usepackage{amsmath, graphicx, tabularx}
\usepackage{amsthm}
\usepackage{bbm}
\usepackage{array}
\usepackage{subcaption}
\usepackage[numbers]{natbib}
\usepackage{hyperref}

\usepackage{amssymb }
\usepackage{bm}
\makeatletter
\newcommand{\multiline}[1]{ %
  \begin{tabularx}{\dimexpr\linewidth-\ALG@thistlm}[t]{@{}X@{}}
    #1
  \end{tabularx}
}

\usepackage{mathtools}

\usepackage{multirow}

\def\mS{\mathcal{S}}

\def\n{_n}

\def\mA{\mathcal{A}}

\def\k{_{k}}

\def\l{_{l}}
\def\j{^{j}}
\def\bt{{\theta}}
\def\w{{w}}

\def\mS{\mathcal{S}}

\def\mA{\mathcal{A}}

\def\t{_{t}}

\def\j{_j}

\def\s{{s}}

\title{EX-DRL: Hedging Against Heavy Losses with EXtreme Distributional Reinforcement Learning}

\date{}

\author{ \hspace{1mm}Parvin Malekzadeh \\
	  University of Toronto\\
	\texttt{p.malekzadeh@mail.utoronto.ca} \\
	\And
	\hspace{1mm} Zissis Poulos\\
	York University\\
	\And
	\hspace{1mm} Jacky Chen \\
	University of Toronto \\
	\And
	\hspace{1mm} Zeyu Wang \\
		University of Toronto\\
	\And
	\hspace{1mm} Konstantinos N. Plataniotis \\
	University of Toronto
}

\begin{document}
\maketitle

\begin{abstract}
Recent advancements in Distributional Reinforcement Learning (DRL) for modeling loss distributions have shown promise in developing hedging strategies in derivatives markets. A common approach in DRL involves learning the quantiles of loss distributions at specified levels using Quantile Regression (QR). This method is particularly effective in option hedging due to its direct quantile-based risk assessment, such as Value at Risk (VaR) and Conditional Value at Risk (CVaR). However, these risk measures depend on the accurate estimation of extreme quantiles in the loss distribution's tail, which can be imprecise in QR-based DRL due to the rarity and extremity of tail data, as highlighted in the literature. To address this issue, we propose EXtreme DRL (EX-DRL), which enhances extreme quantile prediction by modeling the tail of the loss distribution with a Generalized Pareto Distribution (GPD). This method introduces supplementary data to mitigate the scarcity of extreme quantile observations, thereby improving estimation accuracy through QR. Comprehensive experiments on gamma hedging options demonstrate that EX-DRL improves existing QR-based models by providing more precise estimates of extreme quantiles, thereby improving the computation and reliability of risk metrics for complex financial risk management. The code is available at \href{https://github.com/pmalekzadeh/EX-DRL}{https://github.com/pmalekzadeh/EX-DRL}.
\end{abstract}

\keywords{Distributional Reinforcement Learning \and Generalized Pareto Distribution \and Option Hedging}
\section{Introduction}
Option trading desks, especially in high-volume environments like banks, must engage in frequent (e.g., daily) hedging to manage risk and maintain profitability. Specifically, traders try to minimize the sensitivity of their option portfolios to sharp or broad movements in the underlying stock price, as well as to the stock’s volatility. These sensitivities, known as the `Greeks,' when reduced or eliminated, incur costs and directly affect the expected value and distribution of losses from hedging over time. Ideally, traders aim for minimal expected losses and narrower loss distributions, indicating reduced risk exposure. However, the typical hedging strategies employed are `myopic'—they make decisions based on current conditions without considering future implications, leading to suboptimal risk management, particularly in the presence of trading frictions.
\\
To alleviate this problem, a recent thread of research has proposed an alternative take by exploring the application of Reinforcement Learning (RL) to option hedging~\cite{buehler2019deep, vittori2020option, Buehler2022, Cao2021, murray2022deep}, since RL can efficiently learn long-horizon strategies in a stochastic environment. More recent advancements, as demonstrated in \cite{Cao2023a}, show that Distributional RL (DRL)~\cite{dabney2018distributional, bellemare2017distributional}—which models the entire loss (negative reward) distribution for every potential hedging state and action—can surpass classical RL strategies for European options, which typically focuses on expected loss as a minimization criterion.

Quantile Regression (QR) is typically used in DRL algorithms to approximate the loss distribution. Specifically, the loss is modeled as a quantile distribution, represented by a uniform mixture of Dirac deltas at locations corresponding to quantile values. These quantile values are derived using the QR loss function, which is applied within the framework of the distributional Bellman equation. QR is conceptually straightforward and directly assesses risk measures such as Value-at-Risk (VaR) and Conditional Value-at-Risk (CVaR) through the quantile-based representation it creates. However, as highlighted in several studies \cite{bai2021understanding, pan2020uncertainty, beranger2021estimation}, the accuracy of these quantile value estimates, especially those representing extreme values, relies on observations from the distribution's tails. These observations are infrequent yet carry high severity, making the estimation of extreme quantile values inaccurate and prone to under-coverage bias compared to the desired coverage level in practice~\cite{bai2021understanding}. Consequently, sparse data in the extreme tail ends of the distribution render the resulting risk measures unreliable.

To enhance the ability of DRL to model the tails of loss distributions effectively while maintaining scalability and fidelity in representing the rest of the distribution, this paper presents a hybrid model. In this model, the tail of the loss distribution is separately modeled using a Generalized Pareto Distribution (GPD), recognized for its efficiency in capturing extreme values in time series analysis~\cite{pan2020uncertainty, beranger2021estimation, singh2013extreme}, while the main body of the distribution remains a quantile distribution. Termed Extreme-DRL (EX-DRL), our proposed approach is compatible with any QR-based DRL method and substantially enhances the estimation of extreme quantiles, which is crucial for hedging portfolios of options. In our experiments, we combine our hybrid model with QR-Distributed Distributional Deterministic Policy Gradients (QR-D4PG), a recently introduced QR-based DRL algorithm. We refer to the proposed algorithm as EX-D4PG and compare its performance to that of QR-D4PG \cite{Cao2023a}. The findings across a wide range of market scenarios are:
\begin{itemize}
\item EX-D4PG learns a better policy by improved estimates of VaR and CVaR, highlighting the agent's enhanced capability to minimize extreme losses and strengthen risk management.
\item EX-D4PG achieves a better Profit and Loss (PnL) distribution with a higher mean and a narrower spread than QR-D4PG.
\item The hedging policy of the EX-D4PG agent demonstrates increased sensitivity to changes in the volatility of the underlying assets, enabling more responsive adjustments.
\end{itemize}

The rest of the paper is organized as follows. Section \ref{related} discusses related work in RL-based option hedging and GPD-based extreme modelling. Section \ref{background} provides background on QR-based DRL and GPDs. Section \ref{sec:problem} explains how DRL can be used for option hedging.  Finally, Section \ref{extreme} describes the proposed EX-DRL framework in detail, and Section \ref{results} presents empirical results.

\section{Related Work} \label{related}
The concept of using RL as an effective policy optimization framework for option hedging dates back to 2009 \cite{pmlr}. However, it was not until more recently that a broader array of RL-based hedging strategies began to emerge \cite{halperin2020qlbs, kolm, buehler2019deep, godin, Cao2021, vittori2020option, murray2022deep, gao2023deeper, daluiso2023cva}, leveraging advances in deep learning and modern RL algorithms.
In terms of the optimization mechanism, these methods vary: some \cite{halperin2020qlbs} adopt value-based algorithms that directly optimize the expected loss, others \cite{vittori2020option, daluiso2023cva} directly optimize the hedging policy using policy-based methods, while the majority \cite{kolm, Cao2021, murray2022deep, gao2023deeper} implement actor-critic methods that optimize both the hedging policy and the expected loss. These diverse approaches have solidified RL as a compelling alternative to traditional hedging strategies.
A critical design decision in these methods is the selection of the objective function, which significantly influences the reward function. Common choices include  VaR or a convex risk measure such as CVaR and dynamic expectiles. The selection among these objective functions and optimization mechanisms presents distinct trade-offs in terms of scalability, optimality, and time-consistency.

While RL has brought improvements in several domains within risk management, risk-averse formulations based on DRL have exhibited superior scalability in multi-option hedging and have empirically outperformed traditional Greek-neutral hedging strategies under various volatility scenarios \cite{Cao2023a, sharma2024hedging}. In DRL, the entire loss distribution is estimated using a categorical, normal, or quantile representation combined with QR. Quantile representation provides the flexibility to stochastically adjust the quantile values at desired levels and has been successfully applied towards learning risk-averse policies in domains such as Atari games and sports \cite{ma2021conservative, liu2022uncertainty, ICASSP2024, luo2023alternative}. However, these applications typically do not encounter events of extreme risk, which are prevalent in option hedging. Consequently, while DRL approaches provide a broad framework to incorporate various risk-return measures, they tend to yield inaccuracies in the tails of the return distribution. This, in turn, affects the accuracy of extreme quantile estimates, including VaR and CVaR, which are of paramount importance in financial risk management.

In the context of modeling extreme values, GPD from extreme value theory \cite{pickands1975statistical}, offers a way to describe the asymptotic behavior of the distribution of data extremes. A large body of research has leveraged the benefits of GPD to model extreme events in time series and financial analysis~\cite{peng2023improving, abilasha2022deep, xu2022extreme, he2022risk, beranger2021estimation}. However, its applicability in the context of DRL-based hedging remains under-explored. 
\\
Complementary to our framework is the formulation by \cite{davar2024catastrophic}. Both methods utilize GPDs to model the tails of distributions in DRL. However, in \cite{davar2024catastrophic}, the methodology relies on Monte Carlo simulations to estimate the loss and finite difference approximations to compute gradients, which is a suitable approach for low-dimensional policy spaces. Our method, on the other hand, uses deep neural networks to parameterize the policy and improve it via distributional Bellman updates, making it suitable for multi-option hedging scenarios involving large action spaces (thus high-dimensional policy spaces).
\section{Background} \label{background}
In this section, we provide a brief background on DRL, focusing on QR-based algorithms, and then discuss GPDs before turning to our specific use case of hedging derivatives portfolios.
\subsection{Distributional Reinforcement Learning}
We consider the problem of sequential decision-making of an agent acting in an environment over discrete time steps $t$ within a time horizon $T$, i.e., $t \in \{1, 2, ..., T\}$. We formulate this as a Markov Decision Process (MDP) specified by $(\mS,\mA, P_{S}, R,  \gamma)$, where $\mS$ is the state space, $\mA$ is the action space, $ P_{S}(.|s,a)$ is the Markovian transition function that assigns to each state-action pair $(s,a) \in \mS \times \mA$ the probability of reaching the next
state $\s' \in \mS$, $R(s,a)$ is the reward function, and $0<\gamma <1$ denotes the discount factor. The policy of an agent is formulated as $a=\pi(s)$, which maps state $s$ to action $a$ and we assume to be deterministic.

In DRL, the goal is to learn the distribution of the random variable return $G^{\pi}(\s, a)=\sum_{t=1}^{T}  \gamma^{t-1} R(s\t,a\t)$, where $s_1=s$, $a_1=a$, $s\t \sim P_S(.|s_{t-1},a_{t-1})$, and $a\t = \pi(s\t)$. In this paper, we use $Z^{\pi}$ to denote the distribution of the return $G^{\pi}$. It has been shown that $Z^{\pi}$ can be updated using the distributional Bellman equation~\cite{bellemare2017distributional} as:
{
\begin{equation}
Z^{\pi}(\s, a) = T^{\pi} Z(s,a) := R(s,a) + \gamma Z^{\pi}(s', a'), \label{Eq:Bellman_dis}
\end{equation}}  \par  
\noindent{where} $ {s' \!\!   \sim \! P_{S}(.| \s, a)} $, $ a' \! \! = \! \pi(s')$, and $R=R(s,a)$.\footnote{In the rest of the paper, we simplify notation by using $R$ instead of $R(s,a)$.}

\vspace{.1in} \textbf{QR-based DRL:} One of the popular methods to model and learn the return distribution $Z^{\pi}(\s, a)$ in DRL is the QR framework, which was first proposed by~\cite{dabney2018distributional}. QR methods approximate $Z^{\pi}(\s, a)$ using a discrete set of $N$ quantile values $\{ \bt^{\tau\n}_w(s,a)\}_{n=0}^{N-1} $, parameterized with $w$, at specific quantile levels $\{\tau\n = \frac{n}{N}\}_{n=0}^{N-1}$:
\begin{equation}
\!\! Z_{w}(\s, a) = \sum_{n=0}^{N-1} \frac{1}{N} \delta_{\bt_w^{\tau\n}(s,a)}, \label{Eq:return_distribution}
\end{equation} 
where $\delta_{\bt_w^{\tau\n}(s,a)}$ is the Dirac delta function located at $\bt_w^{\tau\n}(s,a)$.\footnote{Throughout this paper, we use $\delta_{x}$ to denote a Dirac delta function located at $x$.} \\
QR-based DRL learns the parameters \( w \) by minimizing the QR loss:
\begin{equation}
\! \! \! \! \! \mathcal{L}^{\text{QR}}(w)  = \sum_{n=0}^{N-1} \mathbb{E}_{z \sim  \hat{Z}(\s, a)} \left[\rho_{\tau_{n}} \left(z- \bt_w^{\tau_n}(s,a) \right) \right],
\! \!   \label{Eq:QR} 
\end{equation}\normalsize \par  
 \noindent  where $\rho_{\tau\n} (e)= e \times ( \tau\n- \mathbbm{1}_{\{ e<0 \}})$. 
 The QR loss can be minimized stochasticity by sampling $M$ values $\{z_m\}_{m=1}^{M}$ from the target distribution $\hat{Z}(\s, a)$, which serves as the benchmark for $Z^{\pi}(s,a)$ during the learning updates. Given a transition tuple $(s,a, R, s', a')$, DRL methods, including QR, derive the target distribution $\hat{Z}(\s, a)$ based on the distribution of $Z(\s', a')$ using distributional Bellman equation. QR models $Z(\s', a')$ as a mixture of $N$ quantile values $\phi_{w'}^{\tau_n}(s',a')$ parametrized with $w'$:
 \begin{equation}
 Z_{w'}(\s', a') = \sum_{n=0}^{N-1} \frac{1}{N} \delta_{\phi_{w'}^{\tau_n}(s',a')}.
 \end{equation} 
Using the distributional Bellman equation, the target distribution \( \hat{Z}(s, a) \) in QR approaches is thus computed based on \( w' \) as
\begin{equation}
\hat{Z}_{w'}(\s, a) = R +  \gamma   Z_{w'}(\s', a') = \sum_{n=0}^{N-1} \frac{1}{N} \delta_{y_n},
 \end{equation}
 where \( y_n = R + \gamma \phi_{w'}^{\tau_n}(s', a') \). Thus, sampling \( M \) values \( \{z_m\}_{m=1}^{M} \) from \( \hat{Z}_{w'}(s, a) \) corresponds to sampling from \( \{y_n\}_{n=1}^{N} \). After updating \( w \) via a gradient step on \( \mathcal{L}^{\text{QR}}(w) \), the parameter \( w' \) defining the target distribution is updated with the new value of \( w \). Doing so ensures that the improvements made to the model's understanding of \( Z_{w}(s, a) \) are carried forward to the target distribution \( \hat{Z}_{w'}(s, a) \).

QR algorithms were initially applied in value-based DRL algorithms~\cite{dabney2018distributional, iqn}, where only the return distribution is learned via the distributional Bellman equation, and actions are selected based on this learned distribution. However, value-based methods are limited to discrete state and action spaces and cannot be applied to continuous spaces. Recently, the use of QR in actor-critic DRL methods~\cite{Cao2023a, sharma2024hedging} has been proposed, where both the return distribution and the policy are parameterized and learned, making it applicable to continuous state and action spaces.
\subsection{Generalized Pareto Distribution}
Originating from the peaks over thresholds approach in extreme value theory~\cite{pickands1975statistical}, the GPD is commonly employed to model the right-tail probabilities of various distributions. However, it can also be adapted to analyze extremely low values or the left tail, which is our focus here. Below, we detail the adaptation of the GPD for modeling the left tail, accompanied by a formal specification:
\begin{definition} 
Let $X_1, X_2, ... $ be independent random variables with Cumulative Distribution Function (CDF) $F$. Given a small threshold $u$, the CDF of threshold exceedance  $u-X$ conditioned on $X<u$ follows a GPD distribution~\cite{pickands1975statistical},
\begin{equation}
p(u-X \leq x | X < u) = H(x),
\end{equation}
where $H(x)$ is the CDF of the GPD: 
\begin{eqnarray}
H(x) = \begin{cases}
    1-(1+ \frac{\varepsilon x}{\sigma})^{-1 / \varepsilon} & \text{if $ \varepsilon \neq 0$} \\
    1-e^{- \frac{x}{\sigma}} & \text{if $ \varepsilon = 0$},
  \end{cases} \label{Eq:GPD}
\end{eqnarray}
\noindent
where $-\infty < \varepsilon < \infty$ and $0 < \sigma < \infty$ are  the shape and the scale parameters of $GPD$. 
\end{definition} 
When $\varepsilon = 0$, the GPD is light-tailed with an exponential decay tail. If $\varepsilon < 0$,  the distribution has a bounded upper limit, and the tail decays exponentially. When $\varepsilon > 0$, the GPD is considered heavy-tailed~\cite{troop2021bias}. In this case, the distribution exhibits a power-law tail, meaning that the probabilities of observing very small values decay polynomially rather than exponentially. This type of tail behavior is associated with `fat tails,' indicating a higher likelihood of extreme events compared to a normal distribution. Research has demonstrated the efficacy of the heavy-tailed GPD for capturing extreme values theoretically \cite{pan2020uncertainty, beranger2021estimation, singh2013extreme}; therefore, it has been leveraged for measuring extreme events in time series and financial analysis \cite{peng2023improving, abilasha2022deep, xu2022extreme, he2022risk}.
\section{Formulating Hedging with Distributional Reinforcement Learning} \label{sec:problem}
In this section, we explain how we formulate the problem of hedging an options portfolio using DRL. 

The environment, similar to that in \cite{Cao2023a}, simulates a 30-day hedging problem that the trading desk of a bank will have to manage to hedge their options exposure resulting from client orders. 
We assume that the trader rebalances her position at fixed intervals of $d t=1$ day leading up to the maturity of the option at $T=30$ days. Moreover, we set $\gamma =1$ as $T$ is relatively short.

The price of the underlying asset, $S$, is assumed to follow a Brownian motion and risk-neutral behavior governed by:
\begin{equation}
dS = (r - q)S \, dt + \sigma S \, dz,
\end{equation} 
where $dz$ is a Wiener process, and $\sigma$ is the volatility of the underlying asset. The risk-free interest rate $r$ and the dividend yield $q$ are assumed to be constant. Additionally, we assume that the real-world expected return $\mu$ is also constant. The real-world process for the asset price is the same as that given above, with $r - q$ replaced by $\mu$.

The value of the option is given by the Black–Scholes–Merton option pricing model developed by \cite{BlackScholes1973} and \cite{Merton1973}:
\begin{equation}
S_0 N(d_1) e^{-q T} - K e^{-r T} N(d_2), 
\end{equation} 
where
\begin{eqnarray}
d_1 &=& \frac{\ln \left( \frac{S_0}{K} \right) + \left( r - q + \frac{\sigma^2}{2} \right) T}{\sigma \sqrt{T}},
\\
d_2 &=& d_1 - \sigma \sqrt{T},
\end{eqnarray}
and $S_0$ represents  the initial price of the underlying asset, $K$ is the strike price of the option, and $N(\cdot)$ denotes the CDF of the standard normal distribution.

Delta ($\Delta$) is the first partial derivative of the option price with respect to the asset price, and gamma ($\Gamma$) is the second partial derivative with respect to the asset price. Under this model, a natural idea is to regard the value of a European option as a function of $S$ and $\sigma$ for the calculation of these partial derivatives:
\begin{eqnarray}
\Delta &=& N(d_1) e^{-q T},
\\
\text{and\,\,} \Gamma &=& \frac{N'(d_1) e^{-q T}}{S_0 \sigma \sqrt{T}}.
\end{eqnarray}
The delta and gamma of a portfolio are calculated by summing those for the individual options in the portfolio. 

\textbf{State:} The state variable $s\t$ at time step $t$ consists of:
\begin{itemize}
    \item The price of the underlying asset.
    \item The gamma of the portfolio.
    \item The gamma of the at-the-money option used for hedging.
\end{itemize}

\textbf{Action:} The agent is allowed to trade an at-the-money call option once every day to maximize the utility against the objective function. Daily, the agent's action $a_t$ is to determine the proportion of maximum hedging done against the total gamma exposure. In other words, the portfolio gamma after hedging relative to gamma before hedging has to be between 0 and 1. This constraint effectively forces the agent to consider hedging as opposed to speculating. Once the agent determines the gamma hedging ratio for the day, the resulting delta will be rebalanced to 0 automatically. 

\textbf{Reward:} 
Each day, we assume new options orders arrive following a Poisson process with a 50/50 chance of being long or short. Following market convention, the options provide the holder the right to buy or sell 100 units of the underlying assets.
The reward (negative cost) $R(s\t, a\t)$ is therefore
\begin{equation}
R(s\t, a\t) = -\kappa \left| V_t H_t \right| + \left( P_t - P_{t-1} \right),
\end{equation}
where $V\t$ is the value of the option used for hedging at time $t$, $H\t$ is the position taken in the option used for hedging at time $ t$, $\kappa$ is the transaction cost associated with the option used for hedging as a proportion of the value of the option, and $P\t$ and $P_{t-1}$ are the portfolio’s market values at time $t$ and $t-1$, respectively, that have not previously expired. If an option expires at time $t$,  its value at time $t$ is set equal to its intrinsic value $\max(S - K, 0)$.

\textbf{Risk Measure:} Traditionally, financial risk management focuses on minimizing risk measures derived from the loss distribution, particularly managing risks associated with high-cost outcomes in the right tail of the loss distribution. However, to align with DRL standard, we redefine our approach based on the reward (return) distribution $Z^{\pi}(s,a)$ of $G(s,a)= \sum_{t=1}^{T}  \gamma^{t-1} R(s\t,a\t)$, effectively shifting our focus to the left side of $Z^{\pi}(s,a)$ where potential low-return, high-risk scenarios occur.
We adapt VaR and CVaR to this context for a specific quantile level $\alpha \in (0,1)$, denoted as $\text{VaR}{\alpha}$ and $\text{CVaR}{\alpha}$. $\text{VaR}{\alpha}$ reflects the quantile at which the CDF of $Z^{\pi}(s, a)$ reaches $1-\alpha$, and $\text{CVaR}{\alpha}$ calculates the expected returns below $\text{VaR}_{\alpha}$.
 
 QR is particularly useful here as it directly measures VaR and CVaR through the quantile-based representations of $Z_w(s, a)$. Specifically, $\text{VaR}{\alpha}(Z_w(s, a))$ directly corresponds to the quantile value $\bt_w^{1-\alpha}(s, a)$, and $\text{CVaR}{\alpha}$ is calculated as the average of quantile values $\bt_w^{\tau_n}(s, a)$ that are less than or equal to $\text{VaR}{\alpha}(Z_w(s, a))$.

However, estimating VaR and CVaR via QR becomes challenging when $\alpha$ approaches $1$. This difficulty arises because observations for extreme quantile values in the tail of the distribution are scarce in samples from the target distribution (i.e., \( \{y_n\}_{n=1}^{N} \)), leading to volatile estimates of these values (i.e., $\bt_w^{1-\alpha}$ ) and consequently, the risk measures~\cite{bai2021understanding, beranger2021estimation}.
%
Evaluating the probability of extreme events has become a pivotal issue in financial risk management, often requiring high $\alpha$ values (e.g., $0.95, 0.99$).  While increasing the number of quantiles $N$ can theoretically enhance the estimation of extreme quantiles by providing more observations at extreme values, this approach has its challenges. Each additional quantile represents new parameters to estimate, potentially leading to a high-dimensional parameter space. This parameter increase complicates the model, heightens the risk of overfitting, and escalates the computational demands. These complexities highlight the need for more robust methods to handle extreme values in DRL for risk management.

Given the success of the QR framework in risk assessment and the effectiveness of the GPD in modeling distribution tails, the subsequent section will introduce our framework, EX-DRL, for options hedging. This framework leverages GPD in conjunction with QR to accurately capture the extreme quantile values of the return distribution $Z(\s, a)$. 
\section{EX-DRL: EXtreme Distributional Reinforcement Learning } \label{extreme}

EX-DRL enriches samples in the tail regions of the return distribution, by not imposing rigid distributional assumptions on the central part of the target distribution of $\hat{Z}(s,a)$ and focusing instead on the extreme values in its left tail modeled with a heavy-tailed GPD. This provides more robust estimates for extreme quantile values and allows us to obtain quantile values directly using the QR method while enhancing their estimation through extensive sampling from the GPD, without the need to increase the number of quantiles $N$.

To benefit from directly measuring VaR and CVaR from the estimated return distribution, we model it as $Z_{w}(\s, a)$, a mixture of $N$ quantile values $\{ \bt_w^{\tau_n}(s,a)\}_{n=0}^{N-1} $ similar to QR-based DRL, as expressed in Eq.~\eqref{Eq:return_distribution}. However, the target distribution $\hat{Z}(s,a)$ is modeled as a mixture distribution consisting of the tail distribution for extreme values, represented using a heavy-tailed GPD, and an unknown body distribution for non-extreme values modeled as a quantile distribution.

Given that the target distribution $\hat{Z}(s,a)$ is derived from $Z(\s', a')$, it follows that $Z(\s', a')$ must also present a mixture distribution. To facilitate this, we introduce a threshold $u(s',a')$, defining the demarcation between extreme and non-extreme values. This threshold is crucial as it determines the separation point in the distribution $Z(\s', a')$, which is then partitioned into two distinct components:
\begin{enumerate}
 \item \textbf{Tail Component $\bm{Z^{Tail}(\s', a')}$}: Represents the behavior of extreme values below $u(s',a')$. It employs a GPD, denoted as $GPD_{\psi}(s',a')$, where the scale parameter $\sigma_{\psi}(s',a')$ and the shape $\varepsilon_{\psi}(s',a')$ parameter are defined by $\psi$. This GPD is negated and shifted so that its endpoint aligns with the threshold $u(s',a')$:
\begin{eqnarray}
\,\,\,\,\,\,\,\,\,\,\,\, Z^{Tail}_{\psi}(\s', a')= - GPD_{\psi}(s',a') + u(s',a'). \label{Eq:Tail}
\end{eqnarray}
\item \textbf{Body Component $\bm{Z^{Body}(\s', a')}$}: Encompasses the bulk of the distribution greater than or equal to $u(s',a')$ and is modeled as a mixture of $N_{Body} < N$ quantiles $\{\phi_{w'}^{\tau_j}(s',a') \}_{j=0}^{N_{Body-1}}$ greater than or equal to the threshold $u(s',a')$, as 
\begin{eqnarray}
Z^{Body}_{w'}(\s', a')=\sum_{j=0}^{N_{Body}-1} \frac{1}{N} \delta_{\phi^{\tau_j}_{w'}(s',a')}. \label{Eq:next_quantile}
\end{eqnarray}
\end{enumerate}
The relative proportions of $Z^{Body}_{w'}(\s', a')$ and $Z^{Tail}_{\psi}(\s', a')$ are adjusted such that the entire distribution $Z(\s', a')$ sums to unity, integrating both components seamlessly. Thus, the composition of $Z(\s', a')$ is parameterized with $w'$ and $\psi$ and is expressed as 
\begin{eqnarray}
Z_{\psi, w'}(\s', a')= (1-\beta) Z^{Tail}_{\psi}(\s', a') + Z^{Body}_{w'}(\s', a') , \label{Eq:next_state}
\end{eqnarray}
where $\beta \in (0,1)$ defines the proportion of $Z_{\psi, w'}(\s', a')$ considered as the body of the distribution, referred to as the `body proportion.'

\begin{figure}[tp]
    \centering
    \includegraphics[width=1.23\linewidth]{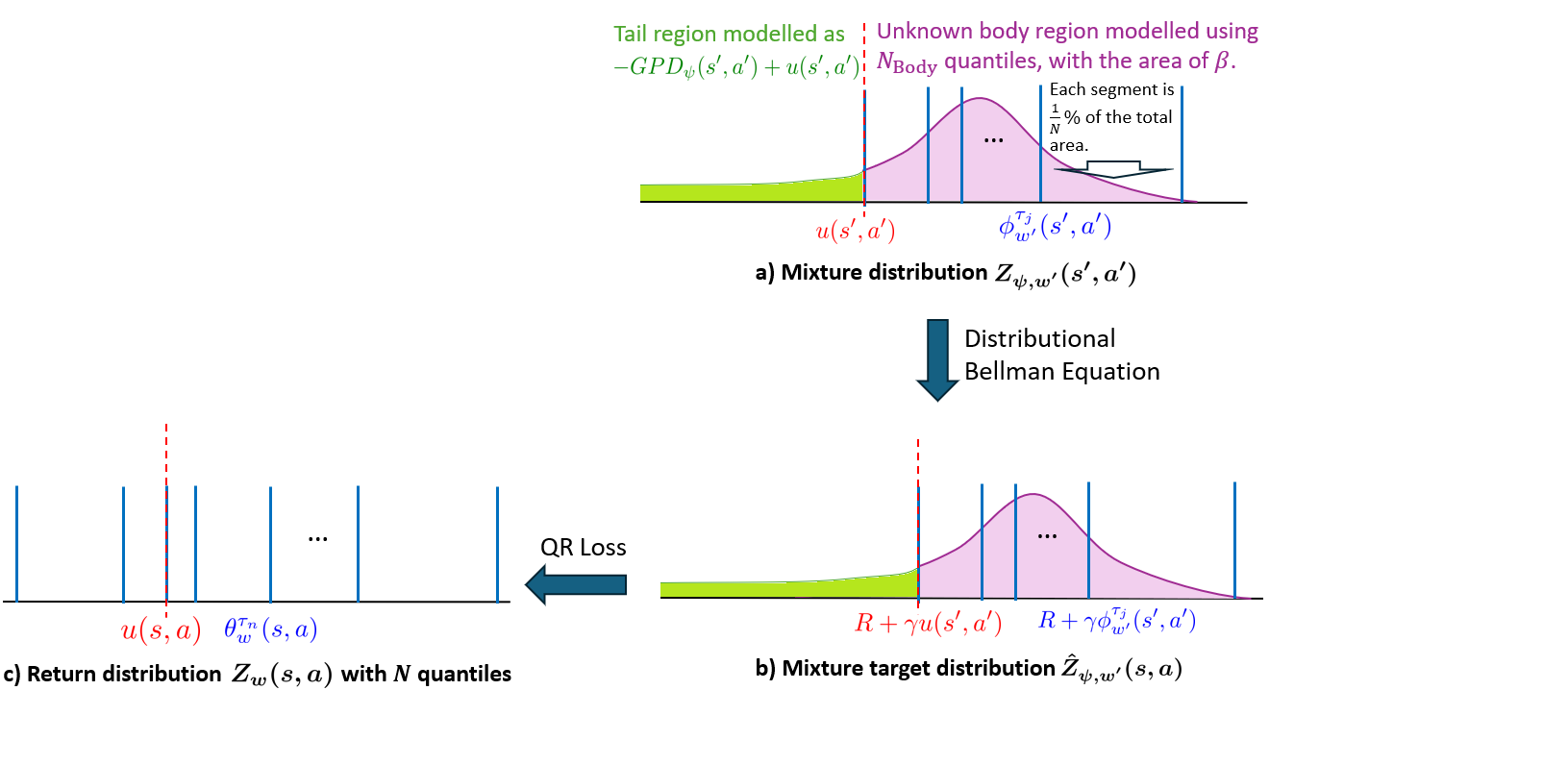}
    \caption{The proposed mixture distribution for the target distribution $\hat{Z}_{\psi, w'}(s,a)$.}
    \label{fig:mixture_model}
\end{figure}

Given this mixture model configuration in Eq.~\eqref{Eq:next_state}, the target distribution $\hat{Z}_{\psi, w'}(s,a)$ is achieved using the distributional Bellman equation as
\begin{eqnarray}
\!\!\!\!\!\!\!\!\! \hat{Z}_{\psi, w'}(s,a) & \!\!\! \!\!\!= \!\!\!\!\!\!& R +  \gamma Z_{\psi, w'}(\s', a') \\
\!\!\!\!\!\!\!\!\! &\!\!\! \!\!\!= \!\!\!\!\!\!&  R +  \gamma  \left( (1-\beta) Z^{Tail}_{\psi}(\s', a') + Z^{Body}_{w'}(\s', a')  \right) \!. \label{Eq:target}
\end{eqnarray}
The proposed mixture model for $Z_{\psi, w'}(\s', a')$ and $\hat{Z}_{\psi, w'}(s,a)$ is illustrated in Figure~\ref{fig:mixture_model}. 
\subsection{Sampling from the target distribution}
We now focus on learning the parameters $w$ of the quantiles $\{\bt^{\tau_n}_w(s,a)\}_{n=0}^{N-1}$ of $Z_w(\s, a)$ using samples from the target distribution $\hat{Z}_{\psi, w'}(s,a)$ defined in Eq.~\eqref{Eq:target}. To achieve this, we first sample from the tail and body components of $Z_{\psi, w'} (\s', a')$ as follows:
\begin{enumerate}
    \item \textbf{Sampling from  $\bm{Z^{Tail}_{\psi}(\s', a')}$:} To better estimate quantiles in the tail of the distribution, we generate a specific number  $M_{\text{Tail}}$ of samples from $Z^{Tail}_{\psi}(\s', a')=  - GPD_{\psi}(s',a') +u(s',a')$.  We achieve this by first generating $M_{\text{Tail}}$ samples from $GPD_{\psi}(s',a')$, then negating each sample and shifting it by $u(s',a')$.
\item \textbf{Sampling from $\bm{Z^{Body}_{w'}(\s', a')}$:}  Given the proportional weight of the body and tail components in $\hat{Z}_{\psi, w'}(s,a)$, the number of samples from  $Z^{Body}_{w'}(\s', a')$ must be $M_{\text{Body}} = \text{round}( \frac{ M_{\text{Tail}} \times \beta}{1-\beta})$. Sampling from $Z^{Body}_{w'}(\s', a')$ corresponds selecting $M_{\text{Body}}$ samples from $\{\phi_{w'}^{\tau_j}(s',a')\}_{j=0}^{N_{Body}-1}$, as specified in Eq.~\eqref{Eq:next_quantile}.
\end{enumerate}
By combining samples from $Z^{Tail}_{\psi}(\s', a')$ and $Z^{Body}_{w'}(\s', a')$ and applying the distributional Bellman equation, we obtain samples $\left\{ \{z\k^{\text{Tail}} \}_{k=1}^{M_{\text{Tail}}}, \{z\l^{\text{Body}} \}_{l=1}^{M_{\text{Body}}} \right\}$ from the target distribution $\hat{Z}_{\psi, w'}(s,a)$.
\subsection{Updating return distribution parameters} \label{subsec:critic_update}
Given the samples $\left\{ \{z\k^{\text{Tail}} \}_{k=1}^{M_{\text{Tail}}}, \{z\l^{\text{Body}} \}_{l=1}^{M_{\text{Body}}} \right\}$, the quantile parameters $w$ of $Z_{w}(\s, a)$ are updated using the QR loss as: 
\begin{eqnarray}
\! \! \! \! \! \mathcal{L}^{\text{QR}}(w) &=& \sum_{n=0}^{N-1} \bigg[ \frac{1}{M_{\text{Body}}} \sum_{l=1}^{M_{\text{Body}}} \rho_{\tau_{n}} \left( z\l^{\text{Body}}- \bt_w^{\tau_n}(s,a) \right) \nonumber \\
&+& \frac{1}{M_{\text{Tail}}} \sum_{k=1}^{M_{\text{Tail}}} \rho_{\tau_{n}} \left( z\k^{\text{Tail}}- \bt^{\tau_n}_w(s,a) \right)\bigg].
\! \!   \label{Eq:EX-QR} 
\end{eqnarray}
The QR loss function integrates contributions from both the body and tail of the return distribution. By explicitly incorporating $M_{\text{Tail}}$ samples from the tail, the model better addresses rare but potentially significant events, which are often under-represented in standard QR-based methods. Thus, this approach leverages the QR method to directly provide quantile values, while the GPD model of the tail enhances the accuracy of these estimates by generating $M_{\text{Tail}}$ samples in the extreme tail segment of the return distribution.
\subsection{Updating target distribution's parameters}
In the previous subsection, we utilized samples from the target distribution $\hat{Z}_{\psi, w'}$ to update the quantile parameters $w$. With these updated parameters $w$  (and thus the updated quantile values $\bt_w^{\tau_n} $), we now focus on updating parameters of the target distribution $\hat{Z}_{\psi, w'}$. This update specifically targets the parameters  $\psi$ of the tail component  $Z^{Tail}_{\psi}$ and the parameters $w'$ of the body component  $Z^{Body}_{\w'}$.  \\
To accomplish this, we consider any tuple $(s'',a''=\pi(s''))$ where $s'' \in \mS$, and proceed as follows:
\begin{enumerate}
    \item \textbf{Updating Parameters of $\bm{{Z}^{Tail}_{\psi}}(s'',a'')$:} As  $Z^{Tail}_{\psi}(s'',a'')$ is modeled using $GPD_{\psi}(s'',a'')$, updating its parameters $\psi$ involves adjusting the parameters of $GPD_{\psi}(s'',a'')$. To accomplish this, we follow these steps:
    \begin{itemize}
    \item Identify the quantiles values $\{ \bt_w^{\tau_i}(s'',a'')\}_{i=0}^{N_\text{Tail}-1}$ among $\{ \bt_w^{\tau_n}(s'',a'')\}_{n=0}^{N-1} $ that are below the threshold $u(s'',a'')$.
    \item Transform these values by negating them and shifting them by $u(s'',a'')$ to ensure they fall within the support range of the GPD.
    \item Apply Maximum Likelihood Estimation (MLE) to these transformed values to update the parameters $\psi$.
    \end{itemize}
\item \textbf{Updating Parameters of $\bm{Z^{Body}_{w'}(s'',a'')}$:}  Consistent with QR-based practices, we update the parameters $w'$ of $Z^{Body}_{w'}(s'',a'')$ using the most recent values $w$.
\end{enumerate}
\subsection{ Selection of threshold $u$} 
Until now, we have assumed the existence of a unique threshold $u(s'',a'')$ for each tuple $(s'',a'')$. 
Depending on the specific application, a fixed proportion $\beta \in (0,1)$ is used to define the segment of the mixed distribution $Z_{\psi, w'}(s'',a'')$ considered as the body, occurring at or above $u(s'',a'')$. Thus, the threshold $u(s'',a'')$ corresponds to the quantile value where the CDF of ${Z}_{\psi, w'}(s'',a'')$ is $1-\beta$, i.e.,  $u(s'',a'')= \phi^{1-\beta}_{w'}(s'',a'')$. This method ensures that $u(s'',a'')$ accurately separates the most extreme values from the body based on the predefined proportion $\beta$.
\subsection{Hedging with EX-DRL}
Our proposed EX-DRL approach is compatible with any QR-based method. As described in Section~\ref{sec:problem}, our option hedging problem employs continuous state and action spaces; thus; we implement EX-DRL within an actor-critic fashion. Here, the return distribution $Z_w(s,a)$ acts as the critic, modeled via a neural network with parameters $w$. Simultaneously, the policy $\pi_\xi(s)$, parameterized by $\xi$, is represented by a separate neural network (the actor). Both actor and critic parameters are iteratively updated until convergence.

Drawing inspiration from the effectiveness of QR-D4PG~\cite{Cao2023a}, a recent actor-critic QR-based DRL method, we propose an extension, EX-D4PG, which replaces the quantile distribution used for the target distribution in QR-D4PG with our mixture model target distribution. Unlike QR-D4PG, EX-D4PG's critic not only reflects the usual dynamics of the return distribution but also emphasizes extreme event modeling. During training, the critic's parameters $Z_w(s,a)$ are updated using gradient descent on the QR loss defined in Eq.~\eqref{Eq:EX-QR}. Concurrently, the actor network adjusts its parameters $\xi$ through gradient ascent to maximize its risk measure, i.e., $\text{VaR}\alpha(Z_w(s,a))$ or $\text{CVaR}\alpha(Z_w(s,a))$.

Compared to QR-D4PG, our method includes an additional neural network, parameterized by $\psi$, which models the scale $\sigma_{\psi}$ and shape $\varepsilon_{\psi}$ parameters of the GPD. As noted in~\cite{troop2021bias}, a GPD with $\varepsilon > 1$ results in an infinite mean, which would imply that the CVaR could also be infinite. Therefore, to effectively adapt to markets with heavy-tailed distributions, our neural network is specifically designed to ensure that $0 < \varepsilon_{\psi} < 1$, maintaining a finite expectation for the CVaR and ensuring stability in estimation under extreme market conditions.
The pseudocode of our proposed EX-D4PG is given in Algorithm~\ref{algo:summary}.
%
\begin{figure}[tbp]
\vspace{-0.15in}
\begin{algorithm}[H]
\caption{\textproc{The proposed EX-D4PG framework.}}
\label{algo:summary}
\begin{algorithmic}[1]
\State {\textbf{Input:}  $N$: number of quantiles, $M_{\text{Tail}}$: number of samples from target distribution's tail, $\beta$: body proportion, $\alpha$: learning rate  }
\State {\textbf{Initialize:}  $w$: critic's parameters, $w'$: target's body parameters, $\psi$: target's tail (i.e., GPD) parameters, $\xi$: actor's parameters   }
\For {$t=0,1, ..., T$}
\State Make transition $(s, a, R ,s', a'=\pi_{\xi}(s'))$
\State {Select threshold}  $u(s',a')=\phi_{w'}^{1-\beta}(s',a')$.
\Statex \hspace{\algorithmicindent}\textbf{Sampling from the target distribution:}
\State \multiline{Sample $\{ x\k\}_{k=1}^{M_{\text{Tail}}}$ from $GPD_{\psi}(s',a')$ }
\State \multiline{Define $ \{ y^{\text{Tail}}\k = -x\k + u(s',a') \}_{k=1}^{M_{\text{Tail}}}$}
\State \multiline{Generate tail samples $\{ z\k^{\text{Tail}} = R + \gamma y^{\text{Tail}}\k    \}_{k=1}^{M_{\text{Tail}}}$}
\State \multiline{Define $M_{\text{Body}} = \text{round}( \frac{ M_{\text{Tail}} \times \beta}{1-\beta})$}
\State \multiline{Sample $\{ y^{\text{Body}}\l \}_{l=1}^{M_{\text{Body}}}$ from $\{ \phi_{w'}^{\tau\j}(s',a') \}_{j=0}^{N_{Body}-1}$ for $\tau\j =\frac{j}{N}+1-\beta$}
\State \multiline{Create  body samples $\{ z^{\text{Body}}\l= R + \gamma y^{\text{Body}}\l \}_{l=1}^{M_{\text{Body}}}$}
\Statex \hspace{\algorithmicindent}\textbf{Update critic's parameters: }
\State \multiline{Compute QR loss $\mathcal{L}^{\text{QR}}(w)$ according to Eq.~\eqref{Eq:EX-QR}. }
\State \multiline{Update $w$ as $w \leftarrow w - \alpha \nabla_w  \mathcal{L}^{\text{QR}}(w)$}
\Statex \hspace{\algorithmicindent}\textbf{Update actor's parameters: }
\State \multiline{$\xi \leftarrow \xi+ \alpha \nabla_{\xi} \, \text{(C)VaR}_\alpha(Z_w(s,\pi_{\xi}(s))) $}
\Statex \hspace{\algorithmicindent}\textbf{Update the target distribution's parameters:}
\State \multiline{Consider tuple $(s'', a''=\pi_{\xi}(s''))$}
\State \multiline{Generate $\{ \bt_w^{\tau_i}(s'',a'')\}_{i=0}^{N_\text{Tail}-1}$ for $\tau_i =\frac{i}{N}$  }
\State \multiline{Define $L(\psi)$ as the log-likelihood of $GPD_{\psi}(s'',a'')$  \\
fitted to $\{\bt_w^{\tau_i}(s'',a'')\}_{i=0}^{N_{\text{Tail}}-1}$}
\State \multiline{ Update target's tail (GPD) parameters as $\psi \leftarrow \psi+ \alpha \nabla_{\psi}L({\psi})$ }
\State \multiline{Update target's body parameters as $w' \leftarrow w$ }
\EndFor
\end{algorithmic}
\end{algorithm}
\vspace{-0.2in}
\end{figure}
\section{Experimental Results} \label{results}
To evaluate the performance of EX-D4PG in capturing extreme losses, we conducted a series of hedging experiments and compared the results with those of QR-D4PG~\cite{Cao2023a}.
%
\subsubsection*{\textbf{Setup}}
 The experimental parameters mirror those used in the gamma hedging experiments described in~\cite{Cao2023a}, enabling direct comparisons between the EX-D4PG and QR-D4PG agents. In this setup, client orders arrive according to a Poisson process at an intensity of 1.0 option per day. Each order typically involves a 60-day option on 100 units of the underlying asset, with an equal likelihood of being a long or short position. We assume transaction costs at 1\% of the option price. Hedging is conducted using a 30-day at-the-money option.  The initial stock price is set at \$10, with an annual volatility of 0.3. We evaluated the performance across 5000 test scenarios. The hedge ratio is calculated as the reduction in gamma exposure post-hedging relative to pre-hedging: 
\begin{equation}
\text{Gamma Hedge Ratio} = 1 - \frac{\sum \text{sign}(T) T^+}{\sum \text{sign}(T) |T|}. \label{Eq:GHR}
\end{equation}
\\
A higher gamma hedge ratio indicates that the agent is more likely to hedge away all its gamma exposure, and a gamma hedge ratio of 1 means the agent always fully hedges its gamma exposure.
To determine the value of the body proportion $\beta$ for our mixed distribution model, we tested values $\{0.94, 0.95, 0.96, 0.99\}$ across various volatility levels. The value $\beta=0.95$ was chosen for its superior performance across several test scenarios.
\subsubsection*{\textbf{Results}}
\begin{figure}[tp]
    \centering
    \includegraphics[width=0.495\linewidth]{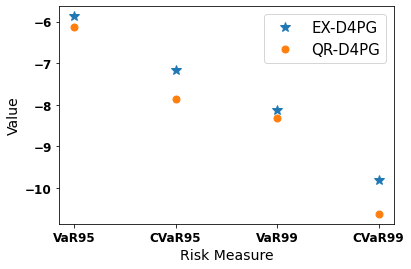}
    \hfill \hfill 
    \includegraphics[width=0.495\linewidth]{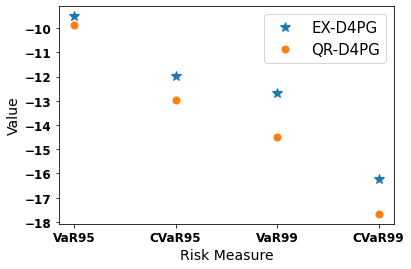}
            \vspace{-0.1in}
    \caption{VaR and CVaR values when volatility = 0.3 (left) and volatility = 0.5 (right).}
    \label{fig:diff_objectives}
            \vspace{-0.15in}
\end{figure}

\begin{figure}[tp]
    \vspace{-0.1in}
    \centering
    \begin{subfigure}[b]{0.6\linewidth}
        \includegraphics[width=\linewidth]{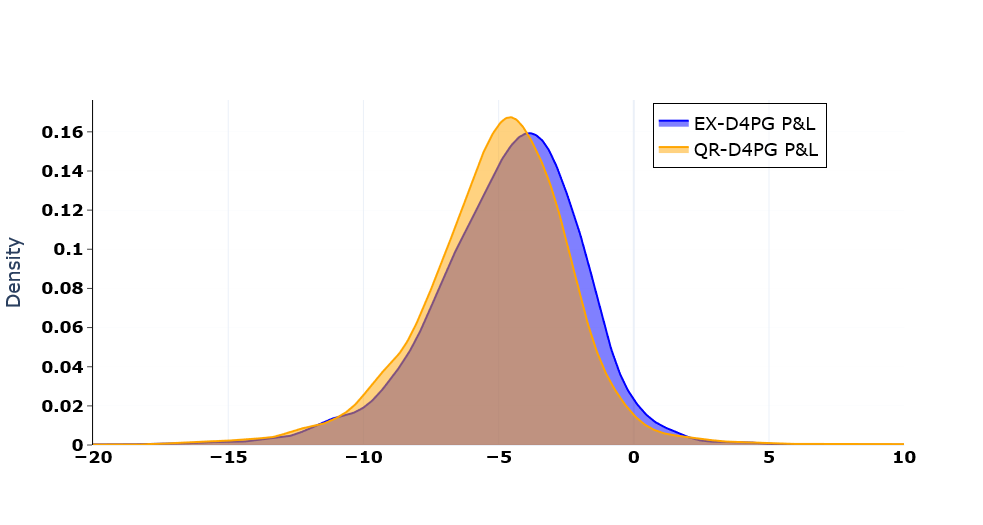}
            \vspace{-0.3in}
        \caption{Risk measure = VaR95}
        \label{subfig:var95}
    \end{subfigure}
    \\
    \begin{subfigure}[b]{0.6\linewidth}
        \includegraphics[width=\linewidth]{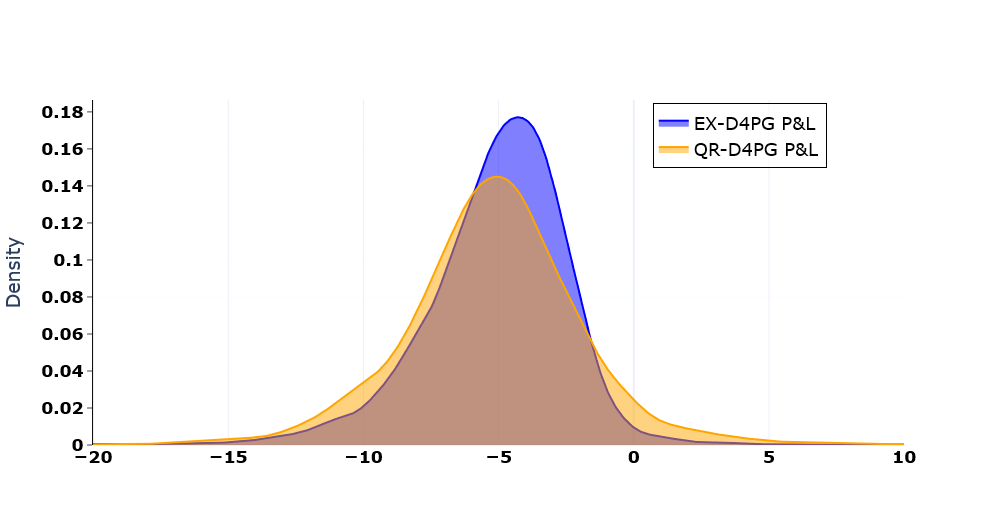}
            \vspace{-0.3in}
        \caption{Risk measure = CVaR95}
        \label{subfig:cvar95}
    \end{subfigure}
    \\
    \begin{subfigure}[b]{0.6\linewidth}
        \includegraphics[width=\linewidth]{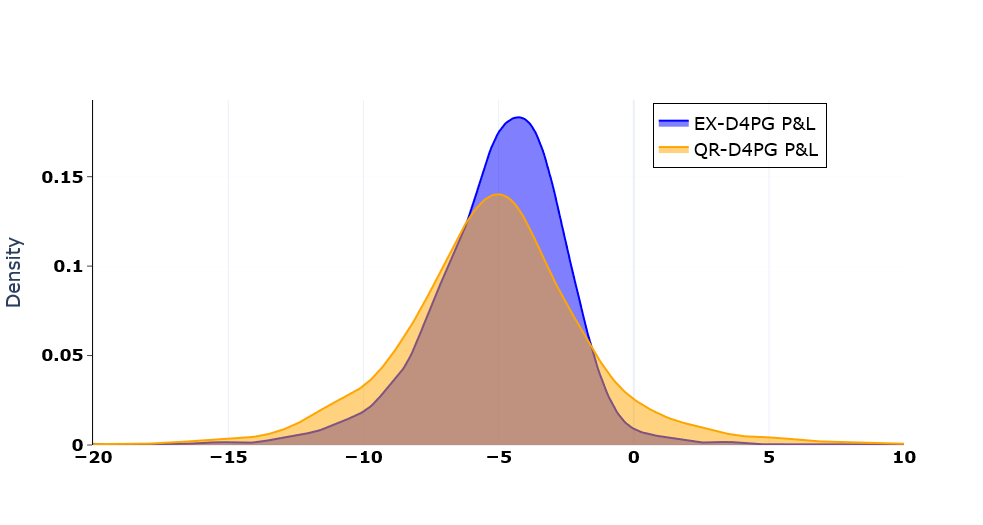}
            \vspace{-0.3in}
        \caption{Risk measure = VaR99}
        \label{subfig:var99}
    \end{subfigure}
    \\
    \begin{subfigure}[b]{0.6\linewidth}
        \includegraphics[width=\linewidth]{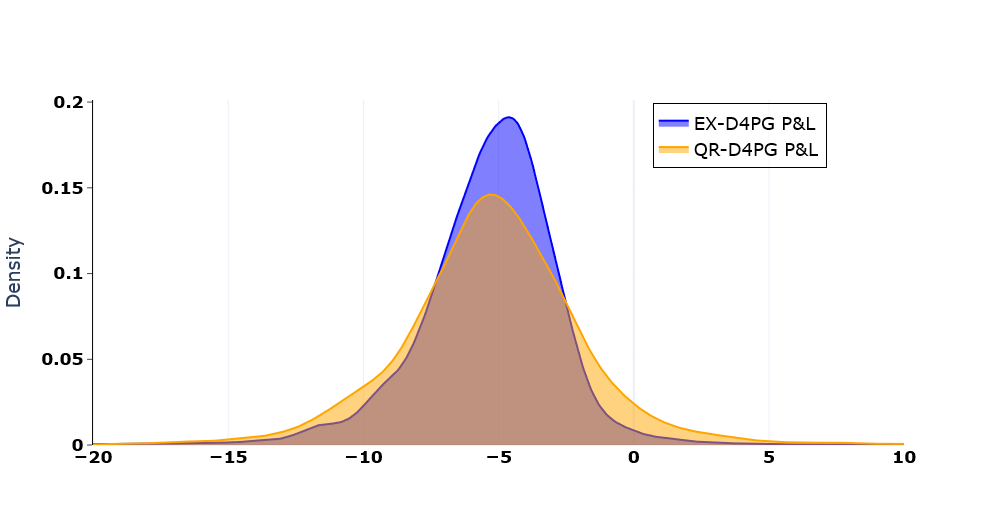}
            \vspace{-0.3in}
        \caption{Risk measure = CVaR99}
        \label{subfig:cvar99}
    \end{subfigure}
    \caption{PnL distributions for different risk measures when volatility  = 0.5.}
    \label{fig:PnL}
\end{figure}

\begin{figure}[tp]
    \centering
    \includegraphics[width=0.49\linewidth]{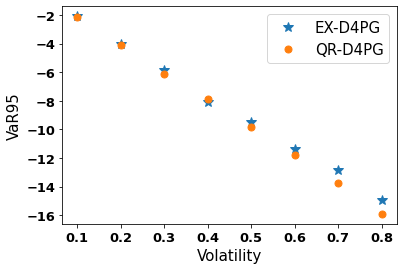}
    \hfill \hfill 
    \includegraphics[width=0.49\linewidth]{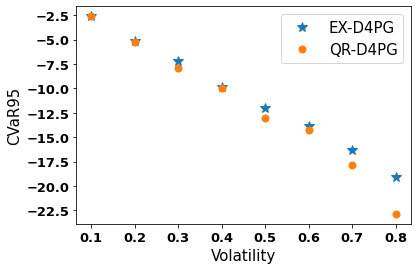}
    \caption{Comparison of VaR95 (left) and CVaR95 (right) at different volatility levels.}
    \label{fig:var-and-cvar95}
\end{figure}

\begin{figure}[tp]
    \centering
    \includegraphics[width=0.48\linewidth]{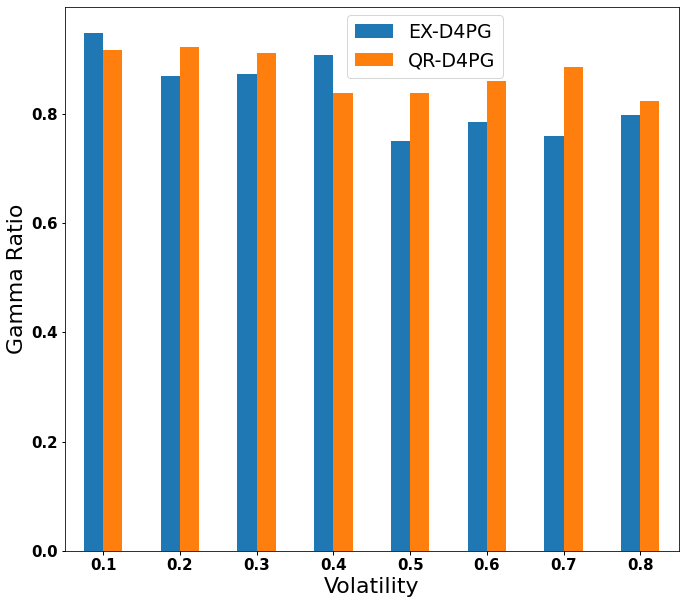}
    \hfill \hspace{.01in}
        \includegraphics[width=0.48\linewidth]{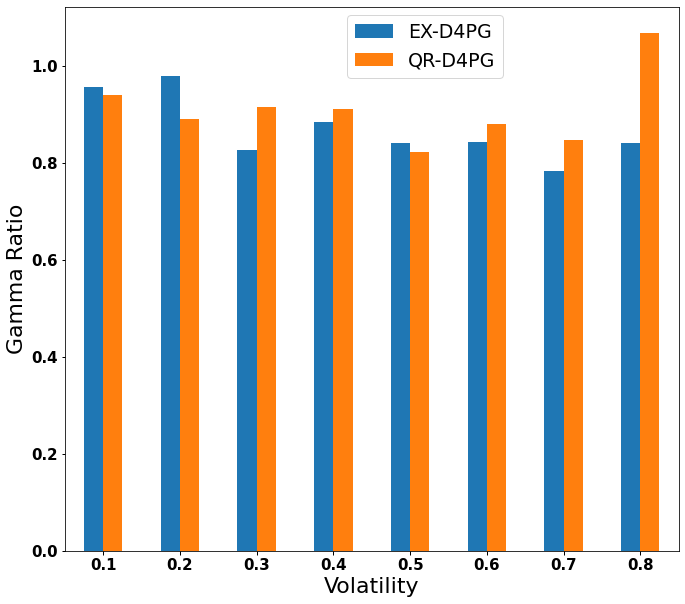}
    \caption{Comparison of gamma hedge ratio using VaR95 (left) and CVaR95 (right) at various volatility levels.}
    \label{fig:Gamma-hedging-ratio}
\end{figure}

Figure~\ref{fig:diff_objectives} compares the performance of QR-D4PG and EX-D4PG at volatility levels of $0.3$ and $0.5$ for the underlying asset. The figure shows that the EX-D4PG outperforms QR-D4PG across all evaluated  risk measures, i.e., $\text{VaR}\alpha$ or $\text{CVaR}\alpha$ for $\alpha=\{0.95, 0.99 \}$. 
\\
Further analysis of the PnL distribution, as depicted in Figure~\ref{fig:PnL}, is supported by the corresponding statistics in Table~\ref{tab:PnL}. These results reveal that the EX-D4PG model incurs fewer extreme losses compared to QR-D4PG. Additionally, the PnL distribution for EX-D4PG  is more on the positive side of the graph, indicating improved mean performance and exhibiting a narrower spread of results, as evidenced by reduced standard deviations.
\begin{table}[bp]
\caption{PnL distribution statistics for different risk measures when volatility = 0.5.}
\centering
\setlength{\arrayrulewidth}{.1mm} 
\setlength{\tabcolsep}{5pt} 
\renewcommand{\arraystretch}{1.} 
\begin{tabular}{|l|c|c|c|}
\hline
\rowcolor{gray!80} \textcolor{white} {\textbf{Risk Measure}} & \textcolor{white}{\textbf{Metric}} &\textcolor{white} {\textbf{EX-D4PG}} & \textcolor{white}{\textbf{QR-D4PG}} \\
\rowcolor{gray!25} \textbf{VaR95} & \multicolumn{3}{c|}{\cellcolor{gray!25}} \\
& Mean  & -4.67 & -5.10 \\
 & Standard Deviation & 2.84 & 2.94 \\
& VaR95  & -9.51 & -9.86 \\
\rowcolor{gray!25} \textbf{CVaR95} & \multicolumn{3}{c|}{\cellcolor{gray!25}} \\
 & Mean & -5.07 & -5.22 \\
 & Standard Deviation & 2.72 & 3.44 \\
 & VaR95 & -9.68 & -10.66 \\
\rowcolor{gray!25} \textbf{VaR99} & \multicolumn{3}{c|}{\cellcolor{gray!25}} \\
 & Mean & -5.00 & -5.08 \\
 & Standard Deviation & 2.62 & 3.82 \\
 & VaR99 & -12.66 & -14.47 \\
\rowcolor{gray!25} \textbf{CVaR99} & \multicolumn{3}{c|}{\cellcolor{gray!25}} \\
 & Mean & -5.35 & -5.25 \\
 & Standard Deviation & 2.65 & 3.50 \\
 & VaR99 & -13.00 & -14.58 \\
\hline
\end{tabular}
\label{tab:PnL}
\end{table}
\\
Moreover, we tested the performance of both RL agents under varying volatility conditions, ranging from $0.1$ to $0.8$, using VaR95 and CVaR95 as the risk measures. The results are shown in Figure~\ref{fig:var-and-cvar95}. The EX-D4PG agent performs comparably to the QR-D4PG at low to medium volatility levels. However, as volatility increases, EX-D4PG demonstrates markedly better performance, underscoring the effectiveness of explicitly modeling the tail distribution in enhancing the agent’s performance under extreme conditions. \\
Finally, Figure~\ref{fig:Gamma-hedging-ratio} analyzes the hedging behavior of the RL agents by comparing the gamma hedge ratio, as defined in Eq.~\eqref{Eq:GHR}. The results show that while both agents generally under-hedge their gamma exposure (indicated by a gamma hedge ratio of less than 1), the EX-D4PG agent demonstrates more adaptive hedging behavior. It adjusts its hedge ratio downward as volatility increases, which helps to reduce hedging costs and improve overall financial performance.

It is important to note from Figure~\ref{fig:Gamma-hedging-ratio} that QR-D4PG slightly outperforms EX-D4PG at volatility levels of 0.1 and 0.4. This is due to using a fixed value of $\beta=0.95$ across all levels to balance model performance with generalizability. Optimizing $\beta$ individually for each level could enhance EX-D4PG's performance, consistently surpassing QR-D4PG. Due to space constraints only outcomes for a fixed $\beta$ are shown.
\section{Conclusion and Future Work}
In this paper, we introduced EX-DRL, a novel DRL algorithm designed to enhance the estimation of extreme quantiles in option hedging. EX-DRL employs a GPD to model the tail of the target distribution and is compatible with QR-based DRL methods, thereby enabling direct estimation of risk measures such as VaR and CVaR. For experimental validation, we integrated the GPD-based tail modeling within a recently proposed QR-based method, QR-D4PG, and termed this integration EX-D4PG. The results confirmed that EX-D4PG outperforms QR-D4PG in precision in extreme quantile predictions, thus providing a more reliable measure of extreme losses.

While our approach currently employs MLE for parameter estimation of the GPD, recent studies~\cite{troop2021bias, huang2019generalized}  suggested more advanced estimation techniques that could potentially enhance parameter accuracy. As a direction for future research, integrating these advanced estimation methods into our EX-DRL framework could further refine the accuracy of extreme quantile estimation.

\bibliographystyle{abbrvnat}
\bibliography{EX-DRL}
\end{document}